\documentclass[fleqn,usenatbib]{mnras}

\usepackage[T1]{fontenc}
\usepackage{ae,aecompl}

\usepackage{graphicx}	% Including figure files
\usepackage{amsmath}	% Advanced maths commands
\usepackage{amssymb}	% Extra maths symbols
\usepackage{txfonts}

\def\gsim{\mathrel{\hbox{\rlap{\hbox{\lower4pt\hbox{$\sim$}}}\hbox{$>$}}}}
\def\lsim{\mathrel{\hbox{\rlap{\hbox{\lower4pt\hbox{$\sim$}}}\hbox{$<$}}}}
\def\Msun{\hbox{$\rm\thinspace M_{\odot}$}}

\title[False periodicities in quasar surveys]{False periodicities in quasar time-domain surveys }

\author[S. Vaughan et al.]{\parbox[h]{\textwidth}{
S. Vaughan,$^{1}$\thanks{E-mail: simon.vaughan@le.ac.uk}
P. Uttley,$^{2}$
A. G. Markowitz,$^{3}$
D. Huppenkothen,$^{4}$
M. J. Middleton,$^{5}$
W. N. Alston,$^{5}$
J. D. Scargle,$^{6}$
W. M. Farr$^{7}$}
\vspace*{4pt} \\
$^{1}$University of Leicester, Department of Physics and Astronomy, Leicester, LE1 7RH, UK\\
$^{2}$Anton Pannekoek Institute for Astronomy, University of Amsterdam, Science Park 904, 1098 XH Amsterdam, The Netherlands\\
$^{3}$University of California, San Diego, Center for Astrophysics and Space Sciences, 9500 Gilman Dr., La Jolla, CA 92093-0424, USA\\
$^{4}$Center for Data Science, New York University, 726 Broadway, 7th Floor, New York, NY 10003, USA\\
$^{5}$Institute of Astronomy, Madingley Road, Cambridge, CB3 0HA, UK\\
$^{6}$NASA Ames Research Center, Astrobiology and Space Science Division, Moffett Field, CA 94035, USA\\
$^{7}$School of Physics and Astronomy, University of Birmingham, Birmingham, B15 2TT, UK\\
}

\date{Accepted 8 June 2016. Received 8 June 2016; in original form 25 May 2016}

\pubyear{2016}

\begin{document}
\label{firstpage}
\pagerange{\pageref{firstpage}--\pageref{lastpage}}
\maketitle

% Abstract of the paper
\begin{abstract}
There have recently been several reports of apparently periodic variations in the light curves of quasars, e.g. PG 1302$-$102 by \citet{Graham2015a}. Any quasar showing periodic oscillations in brightness would be a strong candidate to be a close binary supermassive black hole and, in turn, a candidate for gravitational wave studies. However, normal quasars -- powered by accretion onto a single, supermassive black hole -- usually show stochastic variability over a wide range of timescales. It is therefore important to carefully assess the methods for identifying periodic candidates from among a population dominated by stochastic variability. 
Using a Bayesian analysis of the light curve of PG 1302$-$102, we find that a simple stochastic process is preferred over a sinusoidal variations. We then discuss some of the problems one encounters when searching for rare, strictly periodic signals among a large number of irregularly sampled, stochastic time series, and use simulations of quasar light curves to illustrate these points. From a few thousand simulations of steep spectrum (`red noise') stochastic processes, we find many simulations that display few-cycle periodicity like that seen in PG 1302$-$102. We emphasise the importance of calibrating the false positive rate when the number of targets in a search is very large. 
\end{abstract}

\begin{keywords}
methods: data analysis -- methods: statistical -- surveys, quasars: general -- quasars: supermassive black holes
\end{keywords}

%%%%%%%%%%%%%%%%%%%%%%%%%%%%%%%%%%%%%%%%%%%%%%%%%%

%%%%%%%%%%%%%%%%% BODY OF PAPER %%%%%%%%%%%%%%%%%%

\section{Introduction}
\label{sect:intro}

Detecting and characterising periodic variations in the brightness (and other properties) of astrophysical sources is a cornerstone of observational astronomy. Examples include the discovery of extrasolar planetary systems, using stellar pulsations to establish the cosmological distance scale, and the study of pulsars and interacting binary star systems. Nearly sinusoidal modulations are usually the result of orbital motion or rotation. However, many other astrophysical sources -- notably accreting sources such as interacting binary stars, young stellar objects, and active galactic nuclei (AGN) -- show persistent, random (aperiodic, stochastic, noise) variations in their brightness driven by the complex and turbulent accretion process. See \cite{Vaughan2013} for a brief review of random time series in astronomy. The random variations in AGN can be described as `red noise' -- meaning a random process with a broad power spectrum increasing smoothly in power density to low frequencies (often with an approximately power law shape: $P(f) \propto f^{- \alpha}$, with $\alpha \gsim 1$). 

\defcitealias{Graham2015a}{G15a}
\defcitealias{Graham2015b}{G15b}

\citet[henceforth \citetalias{Graham2015a}]{Graham2015a} reported the detection of periodic modulations (with a period of $5.2$ years) in the optical brightness of the quasar PG 1302$-$102 ($z = 0.278$, $M_V \approx -25.8$, virial mass estimate $M_{\rm BH} \sim 3 \times 10^8$ \Msun) based on $\sim 10$ years of photometric data. This was found during a search of light curves from $243,486$ spectroscopically-confirmed quasars observed with the Catalina Real-time Transient Survey (CRTS; \citealt{Drake2009}). Further details are discussed in \citet[henceforth \citetalias{Graham2015b}]{Graham2015b}. They interpreted their discovery in terms of a short-period binary supermassive black hole system \citep[e.g.][]{Haiman2009b}. Further claims for periodic optical variability in AGN have been made by \citet{Liu2015} and \citet{Zheng2016}. The discovery of short period binary black holes in quasars is of great importance to a number of research areas including accretion physics \citep{Begelman1980},  hierarchical structure formation \citep{Volonteri2003}, and gravitational physics \citep{Haiman2009a}. 

Over the years there have been many reports of periodic or quasi-periodic variations from AGN, spanning the range of AGN types, from radio to gamma-rays, and on timescales from minutes to years. However, this field has a chequered history. Many reports of periodic variations are based on very few observed cycles of the claimed period, and a failure to properly account for the random (red noise) variations which can produce intervals of seemingly periodic behaviour. See \citet{Press1978} for a general discussion of this point, and \citet{Vaughan2006} for some specific examples of periodicity claims drawn just from X-ray observations of nearby AGN\footnote{Arguably the best candidate for quasi-periodic AGN light curve was seen in RE J$1034-396$ \citep{Gierlinski2008}, which showed $\sim 16$ `cycles' in a single, continuous X-ray observation.}. Further observations of the same targets usually fail to show the strictly repeating, coherent oscillations expected from a truly periodic process. As we enter the era of massive time-domain surveys capable of studying $10^5 - 10^7$ targets, it is becoming more important to carefully assess detection procedures in order to understand and control the number of false detections. In this paper we re-examine the case of PG 1302$-$102, and we consider the broader problem of how different stochastic models can make it difficult to distinguish periodic modulation among light curves selected from large time-domain surveys. 

%%%%%%%%%%%%%%%%%%%%%%%%%%%%%%%%%%%%%%%%%%%%%%%%%%
\section{The light curve of PG 1302$-$102}
\label{sect:obs}

\begin{figure}
	\includegraphics[width=\columnwidth]{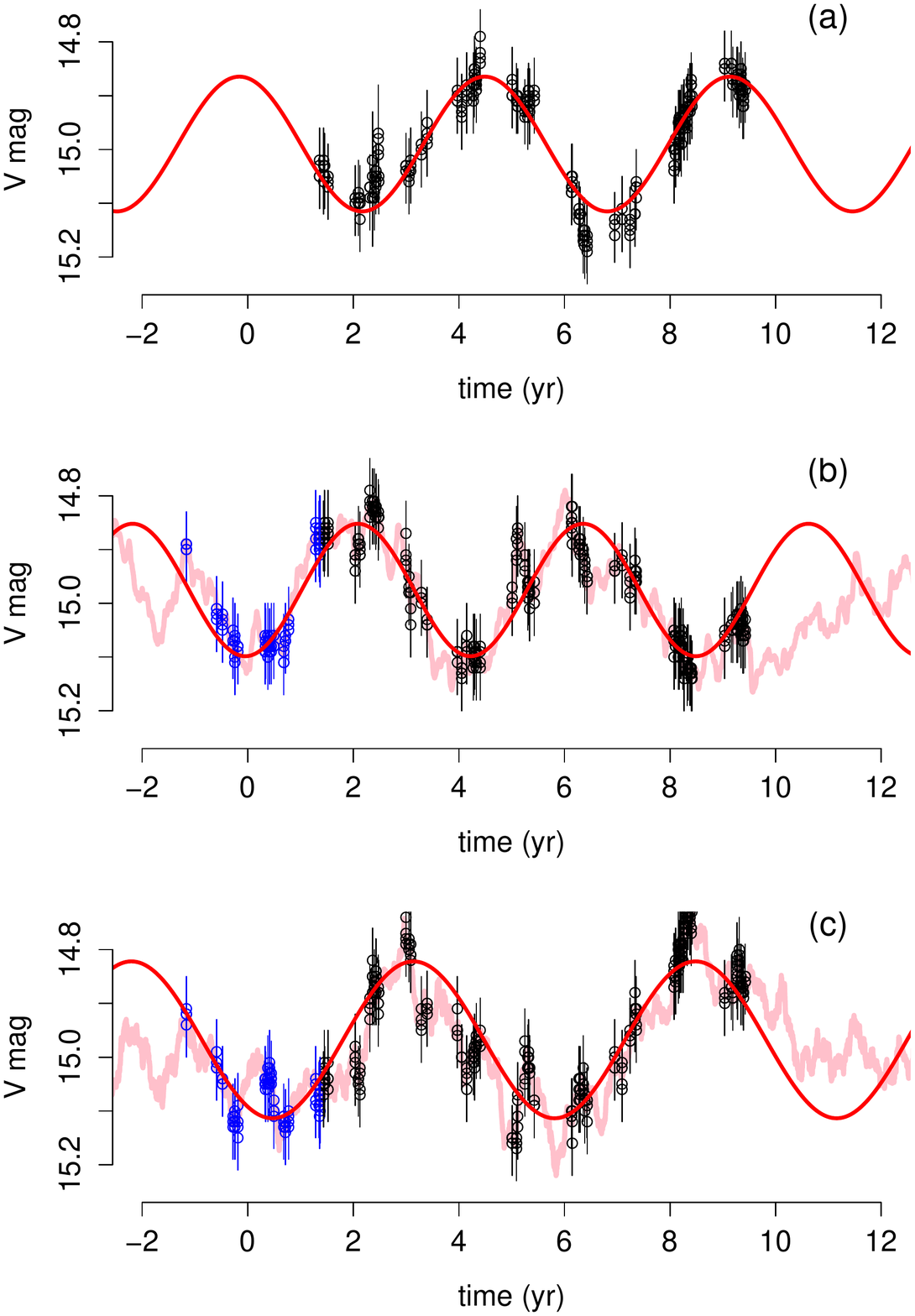}
	\vspace{-1cm}
    \caption{Panel (a) shows the $\approx 8$ years of $V$-band Catalina Real-time Transient Survey (CRTS) data for PG 1302$-$102. Panels (b) and (c) show example simulations of red noise with the same sampling pattern as the CRTS data (black points) plus additional data to simulate three seasons of LINEAR data (blue points). Panels (b) and (c) were generated by random processes with no periodicity present (a bending power law power spectrum, and a damped random walk, respectively). In each case, the continuous, error-free simulation is shown as a pink curve and the sampled data are shown as circles. The red curve shows the best-fitting sinusoid. Examples (b) and (c) were randomly selected from the $100$ best candidates in runs of $100,000$ simulations of each process.}
    \label{fig:examples}
\end{figure}

Figure \ref{fig:examples} (top panel) shows the eight years of CRTS photometric data for PG 1302$-$102 fitted with a sinusoidal model. The data comprise $290$ V-band magnitude estimates with a mean of $\approx 15.0$ mag. The data were taken with two very similar telescopes (CSS and MLS; these provided $234$ and $56$ photometric points, respectively). The sampling pattern is irregular, comprising nine `seasons' each spanning $4-5$ months with gaps of $6-8$ months. Within each season there are $\sim 7$ nights of data, each containing four closely spaced ($\Delta t \sim$ few minutes) photometric measurements. The error bars provided by the CRTS pipeline processing are in this case overestimated by a factor of $\approx 4-5$. This effect can be seen by examining the short timescale variations in the data: the rms variation of the magnitude estimates within groups of nearby data (each group spanning $<20$ days, where intrinsic variability is expected to be weak, and only including groups with $>5$ points) is a factor $\approx 4$ smaller than the error bars\footnote{We have examined CRTS data for other AGN of similar magnitude and find that the photometric error bars are often considerably larger than the short-term scatter in the data.}. 

The data clearly show significant variations, with an rms $\sim 0.1$ mag. We fitted the data (using weighted least squares) with a model comprising a sinusoid plus a constant offset:
\begin{equation}
  V(t) = A_1 \cos(2 \pi f_0 t) + A_2 \sin(2 \pi f_0 t) + C. 
	\label{eqn:sin}
\end{equation}
(This is equivalent to a model $A \sin(2 \pi f_0 t + \phi) + C$ with amplitude given by $A^2 = A_1^2+A_2^2$ and phase $\tan \phi = A_1/A_2$.) The best-fitting amplitude is $(A_1^2 + A_2^2)^{1/2} = 0.125$ mag and the best-fitting (observer frame) period is $t_0 = 1/f_0 = 4.65 \pm 0.06$ yr, slightly different from the $5.16 \pm 0.24$ yr found by \citetalias{Graham2015a}. For fitting their sinusoidal model G15a included additional archival data -- notably LINEAR data \citep{Sesar2011} -- extending the observational baseline. The overall fit statistic is $\chi^2 = 85.7$ for $287$ degrees of freedom, again indicating that the error bars are too large. Comparing this model to a constant gives $\Delta \chi^2 = 741.1$. 

%%%%%%%%%%%%%%%%%%%%%%%%%%%%%%%%%%%%%%%%%%%%%%%%%%
\section{Bayesian model comparison}
\label{sect:bayes}

It is also possible to fit the data using a stochastic model. However, is not meaningful to simply compare the $\chi^2$ values for these fits. When fitting stochastic models to individual time series, the $\chi^2$ fit statistic loses its simple meaning as a diagnostic of the `goodness of fit'. (This is because the variance of the process is itself a parameter to be fitted; the standard $\chi^2$ statistic only makes sense as a likelihood proxy when the variance is fixed. In fact, $\chi^2 \rightarrow 0$ is possible for any sufficiently flexible stochastic process. See also \citealt{Kozlowski2016}). 

In order to compare a periodic model to a stochastic model, we have performed a Bayesian model comparison between the sinusoidal model and a simple stochastic process, the damped random walk model. We first computed the posterior densities for the parameters of each model using Markov Chain Monte Carlo (MCMC) method. We used a method based on the ensemble sampler proposed by \citet{Goodman2010} with $>10^5$ draws based on $100$ `walkers' after removing a `burn-in' period\footnote{The code, called {\tt tonic}, is a pure {\bf R} implementation of the \citet{Goodman2010} sampler, and uses a mixture of `stretch' and `walk' moves to sample the target density. See \url{https://github.com/svdataman/tonic}.}. We then computed the full marginal likelihoods (often called `evidence') of each model using the methods described in \citet{Bailer-Jones2012} and used by \citet{Andrae2013}. 

\begin{figure*}
\hbox{
 	\includegraphics[width=\columnwidth]{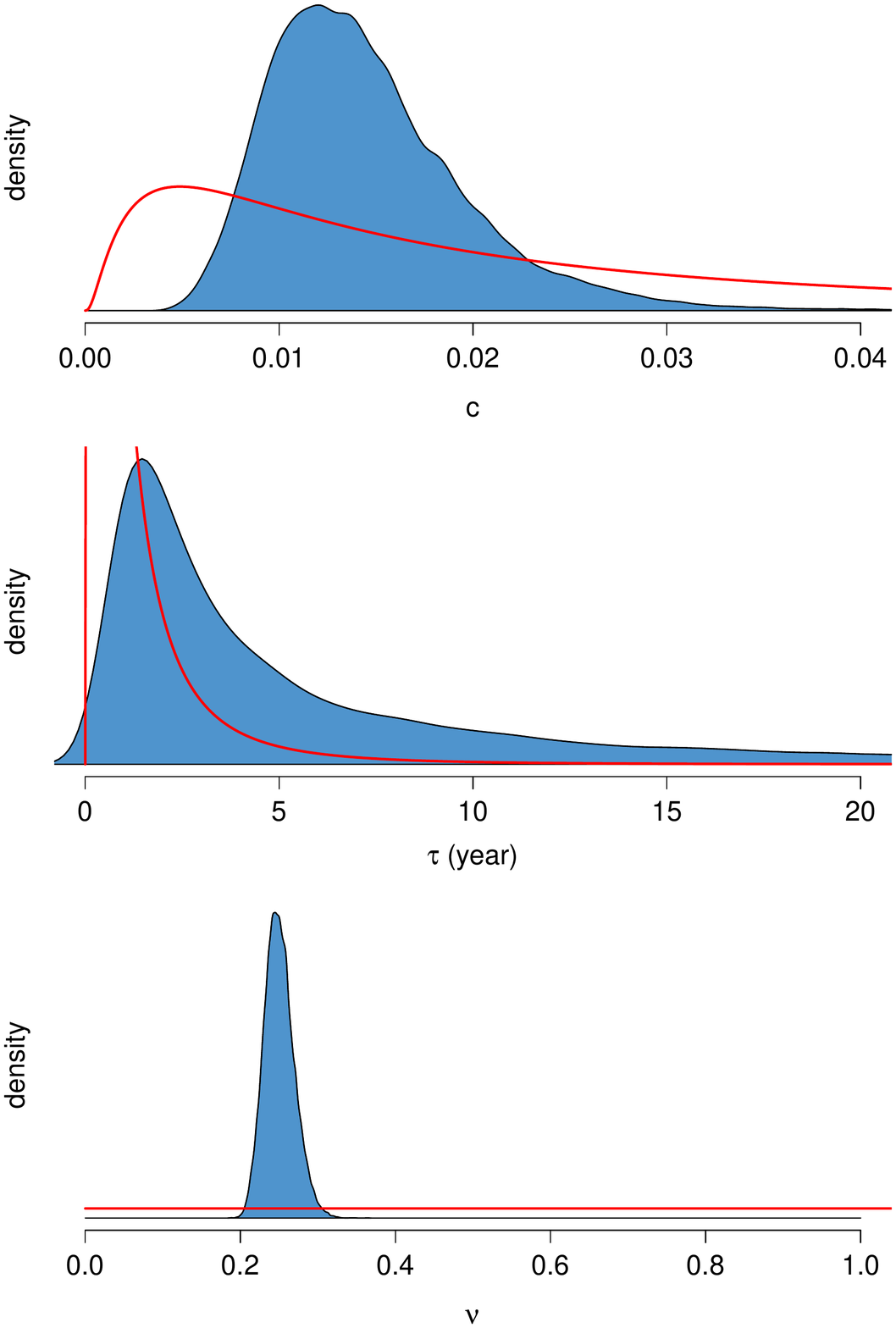}
	\includegraphics[width=\columnwidth]{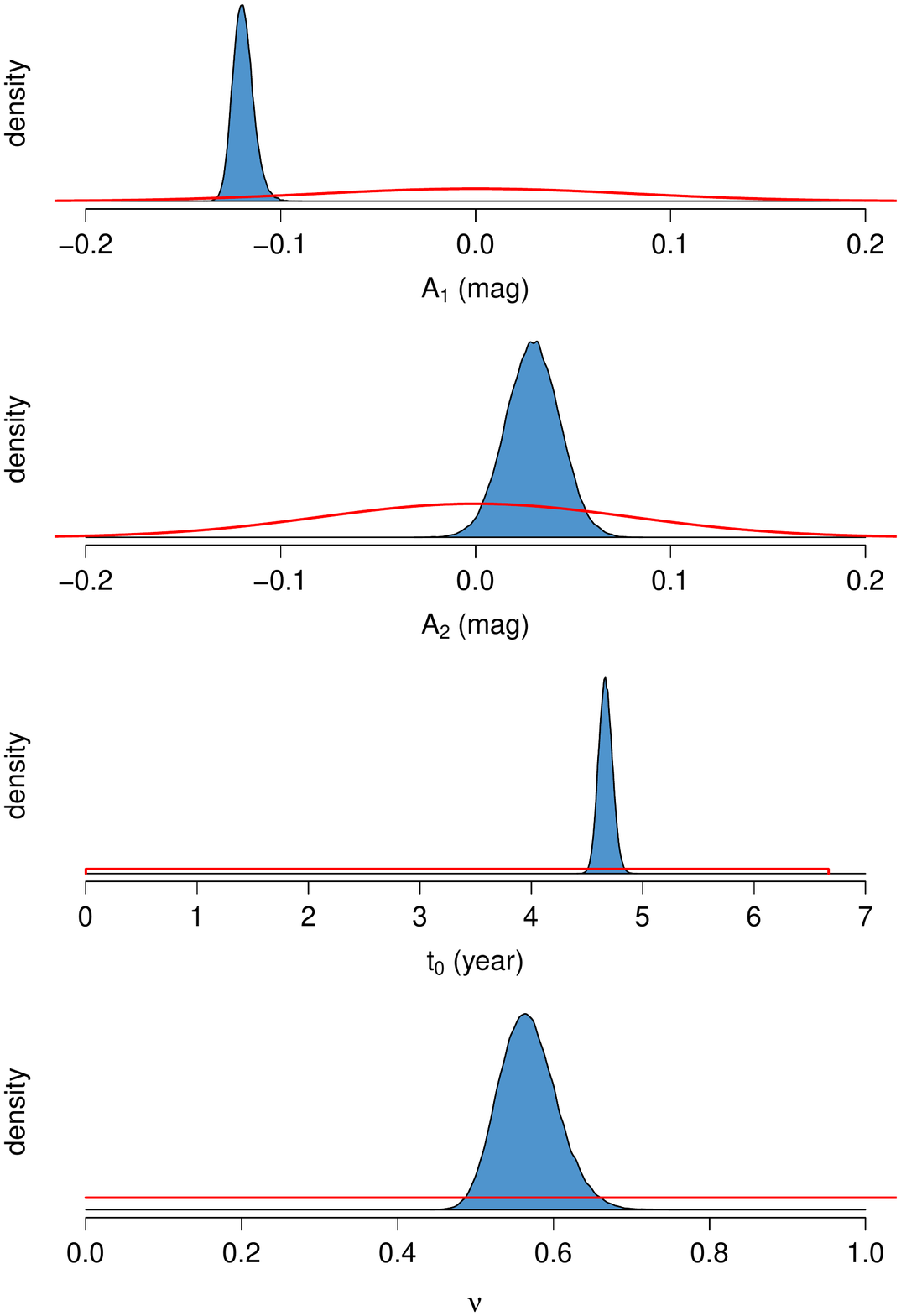}
	}
	\vspace{-0.3cm}
    \caption{Posterior distributions for the parameters of the DRW model (left) and sine model (right) fitted to the CRTS light curve of PG 1302$-$102. The shaded areas show the posterior densities estimated using $10^5$ samples from a Markov Chain Monte Carlo (MCMC) calculation, and the red curves show the prior densities. The priors for the DRW model are based on the sample parameters from large quasar surveys. The priors for the sine model are such that the prior mean variance is similar to the prior mean for the variance of the DRW model, and the period prior is uniform over the range $0-6.67$ yr.}
    \label{fig:mcmc}
\end{figure*}

The damped random walk (DRW) is a stochastic process often used to describe quasar variability from survey data. See \citet{Kelly2009}, \citet{MacLeod2010}, \citet{Andrae2013}, \citet{Zu2013}. It is among the very simplest continuous-time stochastic processes. The DRW has an auto-covariance function $ACV(t) = (c \tau/2) \exp(-t/\tau)$, specified by two parameters, $c$ and $\tau$, which determine the (total) variance and characteristic timescale, respectively. Equivalently, the DRW has a power spectrum (the Fourier transform of the $ACV$) that is flat (power law index $0$) below $f_{\rm bend} = 1/(2 \pi \tau)$, and smoothly bends to a power law with index $-2$ at higher frequencies. We have fitted this model, with three parameters $\{ c, \tau, \nu \}$, to the mean-subtracted PG 1302$-$102 data. The parameter $\nu$ is a scale factor applied to the photometric error bars (see also \citealt{Kelly2014}).
We assigned the following prior densities on these parameters. For $\tau$ and $c$ (which are positive-valued) we assigned lognormal priors, in both cases with $\sigma = 1.15$ (corresponding to $0.5$ dex). The means of each lognormal prior were set based on the geometric mean $\tau$ of $200$ d from quasar samples in \citet{MacLeod2010}, \citet{Kozlowski2010} and \citet{Andrae2013}, and to give $\sigma_T = (c \tau / 2)^{1/2} \sim 0.1$ mag. For $\nu$ we assigned a uniform prior. The posterior and prior densities of these parameters are shown in Figure \ref{fig:mcmc} (left) and summarised in Table \ref{tab}. The posterior shows $\nu$ is low, consistent with the conclusion above (section \ref{sect:obs}) that the error bars are a factor $4$ too large.

\begin{table}
	\centering
	\caption{Summary of prior and posterior distributions.}
	\begin{tabular}{llll} % four columns, alignment for each
		\hline
		parameter & prior & posterior & $90$ per cent\\
		          & description & mode & interval \\
		\hline
		DRW  & & & \\
		\hline
		$c$    & lognormal & $1.2\times10^{-2}$ & $[0.79, 2.4]\times10^{-2}$  \\
		(mag$^2$ yr$^{-1}$) & $\mu = -4.0$, $\sigma=1.15$ & & \\
		$\tau$ & lognormal & $1.5$ & $[0.79, 50.1]$ \\
		(yr) & $\mu = -0.6$, $\sigma=1.15$ & &  \\
		$\nu$  & uniform   & $0.24$ & $[0.22, 0.28]$ \\
		       & min$ = 0$, max$=1.5$ & &  \\
		\hline
		sine & & & \\
		\hline
		$A_1$ & normal & $-0.119$ & $[-0.127, -0.110]$ \\
		(mag)      & $\mu=0$, $\sigma=0.08$ &  & \\
		$A_2$ & normal & $0.031$ & $[0.007, 0.051]$ \\
		(mag)      & $\mu=0$, $\sigma=0.08$ &  & \\
		$t_0$ & uniform & $4.66$ & $[4.57, 4.77]$ \\
		(yr)      &  min$ = 0$, max$=6.67$ &  & \\
		$\nu$ & uniform & $0.56$ & $[0.51, 0.63]$ \\
		      &  min$ = 0$, max$=1.5$ &  & \\
					\hline
					\hline
	\end{tabular}
	\label{tab}
\end{table}

We also fitted a sine model with four parameters $\{ A_1, A_2, t_0, \nu\}$ to the mean-subtracted PG 1302$-$102 data. The model is based on equation \ref{eqn:sin} with  an additional, $\nu$, to rescale the error bars. The two amplitude parameters, $A_1$ and $A_2$, were given zero-mean normal priors (and $\sigma$ chosen to give a prior mean $\sigma_T \sim 0.1$ mag (the same mean prior variance as the DRW model). The period $t_0$ was assigned a uniform distribution (equivalent to using a Pareto prior distribution for the frequency, $f_0$). The $\nu$ parameter was given the same uniform prior as for the DRW model. 

The posterior and prior densities of these parameters are shown in Figure \ref{fig:mcmc} (right) and Table \ref{tab}. 

We then estimated the marginal likelihood of each of these models,
\begin{eqnarray}
  p(D|M) & = & \int p(D,\theta|M) d\theta \nonumber \\
	              & = & \int p(D | \theta, M) p(\theta|M) d\theta, 
\end{eqnarray}
where $\theta$ represents all the parameters of model $M$, $p(\theta|M)$ is their combined prior density, and $p(D|\theta,M)$ is the usual likelihood function given data $D$. The ratio, $B_{12} = p(D|M_1)/p(D|M_2)$, called the `Bayes factor', provides a way to weigh the probabilities of two models, $M_1$ and $M_2$. Here, the two model are the DRW and sine models. 

Marginal likelihoods are usually difficult to compute. We therefore used three methods to calculate $B_{12}$\footnote{The three methods were: simple Monte Carlo integration based on sampling from the prior with $N=10^8$ draws, the Laplace approximation of the posterior distribution, and the $K$-fold cross-validation method described in \citet{Bailer-Jones2012}}. Using all the three methods we found $\log_{10}(B_{12}) > 60$, indicating a very strong preference for the DRW over the sine model. This does not mean the DRW provides an adequate description of the data, only that it is strongly favoured over the sine model. This may at first seem surprising, given the smooth nature of the light curve (top panel of Figure \ref{fig:examples}). However, close inspection reveals that the sine model fails to capture structure in the light curve (such as the different peak magnitudes of the two maxima) that can be modelled by the DRW. This is despite the fact that PG 1302$-$102 was chosen to be among the most periodic from $\sim 250,000$ light curves.

%%%%%%%%%%%%%%%%%%%%%%%%%%%%%%%%%%%%%%%%%%%%%%%%%%
\section{Simulations of red noise}
\label{sect:sims}

In the above analysis we found the simple stochastic model to be strongly preferred over the sinusoidal model, despite the undulating appearance of the light curve of PG 1302$-$102. We next study how often simple stochastic processes produce nearly sinusoidal light curves by producing a number of fake time series with the same sampling pattern as the PG 1302$-$102 CRTS + LINEAR data but generated using
\begin{enumerate}[i]
\item a Gaussian noise process with a steep, bending power law (BPL) power spectrum,
\item a damped random walk (DRW), 
\item a sinusoidal process.
\end{enumerate}
We refer to the three types of process as the BPL, DRW, and sine models, respectively. 

The BPL and DRW are intended to simulate observations of normal (single BH) quasars, but with different assumptions about the typical quasar power spectrum. We initially set the DRW timescale parameter $\tau = 200$ d, the geometric mean of the quasar samples in \citet{MacLeod2010} and \citet{Andrae2013}. 

The BPL model has a steep high frequency power spectrum. This choice of model is motived by the analysis of the high-quality \emph{Kepler} light curves of nearby AGN which showed much steeper power spectra than the standard DRW model allows \citep{Mushotzky2011, Edelson2013, Edelson2014}, with power law indices $\alpha \gsim 3$. Steep high-frequency power spectra are also common in rapid X-ray variability of nearby AGN \citep[e.g.][]{Gonzalez-Martin2012}. It has the form given in section 3.5 of \citet{Summons2007}, with two bend frequencies, $f_{\rm low}$ and $f_{\rm hi}$. We use power law indices of $0$ (below $f_{\rm low}$), $-2$ (between $f_{\rm low}$ and $f_{\rm hi}$), and $-3.5$ (above $f_{\rm hi}$). The frequencies were initially set to $f_{\rm low} = 0.2$ yr$^{-1}$ (timescale $\sim 5$ yr) and $f_{\rm hi} = 7.3$ yr$^{-1}$ (timescale $\sim 50$ d). This is based on the high quality optical power spectrum of Zw 229-15 from \emph{Kepler} data \citep{Edelson2014} with frequencies scaled down by a factor $\sim 10$ as expected for a $M_{\rm BH} \sim \text{few} \times 10^8$ \Msun\ quasar. The sine model simulations are intended to fake data as if from short-period binary quasars, with each simulation having a single modulation period randomly drawn from a uniform distribution over the range $0.2 - 20$ yr (corresponding to a frequency distribution $p(f) \propto f^{-2}$ over the range $0.05 - 5$ yr$^{-1}$). More details of the simulation procedure are given in the Appendix (section \ref{sect:simdets}).

We simulated $100,000$ time series using each of the three processes. We then fitted each with the sinusoidal model. We identified as `periodic candidates' all simulations for which (a) the fit is good ($\chi^2 < dof$) and the improvement in the fit compared to a constant is large ($\Delta \chi^2 > 700$), and (b) the period is in the range $5.38 - 1.25$ yr. The first criterion is used to select data with significant variability that resembles a sinusoidal modulation ($\Delta \chi^2 > 700$ is comparable to that found for PG 1302$-$102 above). The longest allowed period was chosen to match that of \citetalias{Graham2015a} and \citetalias{Graham2015b} who selected only periodic candidates with $>1.5$ cycles in the $9$-year CRTS data. We found it necessary to impose a limit on the shortest allowed periods that is slightly longer than the typical spacing of the CRTS seasons. Allowing shorter periods results in a large number of good fits with periods $\sim1$ yr or shorter, where the quasi-periodic sampling pattern of the CRTS data occasionally aligns with local maxima or minima of the simulation. This is an aliasing effect also discussed in the appendix of \citet{MacLeod2010}. Figure \ref{fig:examples} (b)-(c) show examples of candidate periodicities drawn from simulations of the BPL and DRW processes.

Apart from our choice of the range of accepted periods, the above criteria are not intended to reproduce the period detection methods of \citetalias{Graham2015a} and \citetalias{Graham2015b}, or any other paper. We are simply selecting time series that have a sinusoidal shape (those that give a good match, in a least squares sense, to a sinusoid, and a poor match to a constant model). Any reasonable period detection algorithm should be able to identify the same time series as appearing to be periodic over at least $1.5$ cycles. A more general selection and fitting procedure that allowed for non-sinusoidal periodicities and for additional trends in the data will mostly likely identify additional false periods that were not selected by our fitting, and so our method is conservative. Furthermore, we selected candidates based on fits to $360$ data points spanning $\sim 10.6$ yr (appropriate for a combined CRTS + LINEAR dataset) -- this included more points and a longer baseline than most of the CRTS data used by \citetalias{Graham2015b} -- and so our criteria for selecting periodic variability are in this sense more strict. 

\begin{figure}
	\includegraphics[width=\columnwidth]{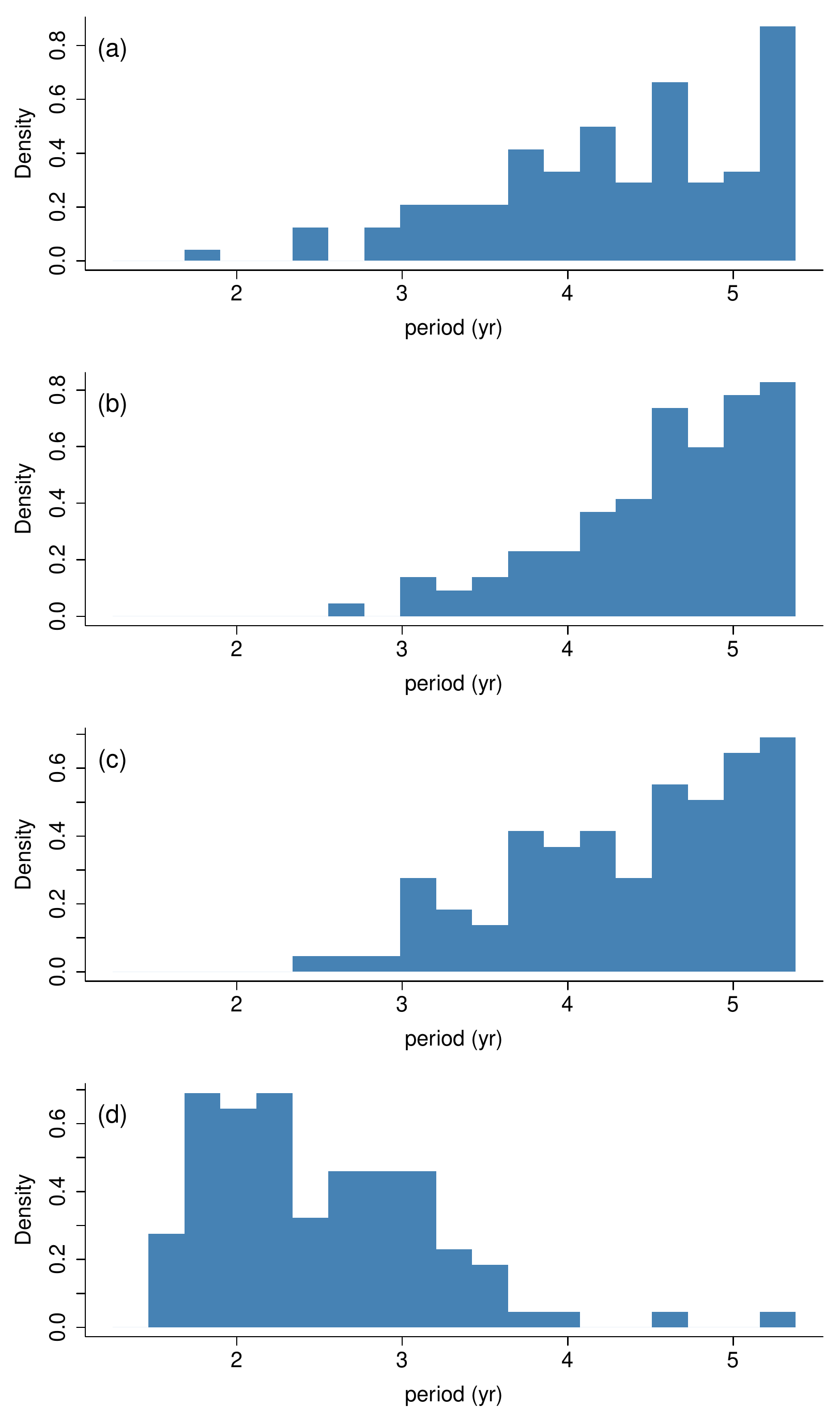}
	\vspace{-0.3cm}
    \caption{Period distributions of periodic candidates. Panel (a) shows the distribution of the $111$ period candidates identified by \citetalias{Graham2015b}. Panels (b)-(d) show the frequencies of the $100$ best candidates from $100,000$ simulations of (b) red noise with BPL spectrum, (c) red noise with a DRW spectrum, (d) sinusoids with uniformly distributed periods. The histograms show the density of periods, i.e. each is normalised such that its total area is unity. There is a clear tendency to find phantom periodicities with long periods ($\sim 5$ yr) while truly sinusoidal signals are most easily recovered at shorter periods (a preference for periods of $\sim 2$ years is expected from sampling theory given the roughly annual spacing of the observing seasons).}
    \label{fig:periods}
\end{figure}

Simulated data meeting our selection criteria were produced with a rate of $\sim 1-2$ per $1,000$ simulations for the DRW and BPL processes, with parameters defined as above. The periods of the fitted sinusoids are long, most are in the range $4.0 - 5.3$ yr (i.e. $1.5 - 2.5$ cycles over the simulated data), and the strongest cases have periods of $\approx 5.3$ yr, always near the lower limit of the allowed range. The distribution of periods is shown in Figure \ref{fig:periods} for the $111$ candidates identified by \citetalias{Graham2015b} from the CRTS data, and from the $100$ strongest period detections in simulations of the BPL, DRW, and sine processes. The steep spectrum (red noise) random processes produce nearly-sinusoidal variations when sampled intermittently, and most of these `phantom periodicities' show only a few cycles, typically less than three\footnote{One way to understand this is in terms of the Fourier decomposition of a realisation of a steep-spectrum stochastic process, sampled at a finite number of discrete times. The observed time series can be decomposed into a finite number of `modes' with different frequencies; modes with lower frequencies have (on average) much higher amplitudes due to the steep power spectrum. But the amplitudes (at a given frequency) fluctuate greatly between different realisations of the same process \citep[e.g.][]{Timmer1995}. With steep spectrum processes it will often be the case that a single low frequency mode dominates the power (variance) of the data, due to random fluctuations.}.

We have repeated the simulation experiments above with different choices for the BPL bend frequencies and the DRW timescale. The rate of phantom periods is highest when $\tau \sim 200 - 400$ d (DRW) or $f_{\rm low} \sim 0.2$ yr$^{-1}$ (BPL). The power spectra that show $f \times P(f)$ peaks ($\sim 1 / 2 \pi \tau$ for the DRW model) near the observable frequency range ($\sim 0.1-1$ yr$^{-1}$) produce time series with strong, smooth variations on the timescales sampled, and are mostly likely to produce phantom periods. \citet{MacLeod2010} and \citet{Andrae2013} found a geometric mean of $\tau \sim 200$ d ($0.55$ yr) from their DRW model fitting to large samples of quasars. This is the right order of magnitude for  phantom periods to be most easily produced in data spread over $\sim$few years. If the DRW spectrum is modified to have a high frequency slope of $3$ (rather than $2$), the rate of phantom periods is increased by a factor of $\sim$few, to $\sim 1$ in $200$ simulations. 

The reduction in the rate of phantom periods with higher/lower DRW characteristic timescale can be understood as follows. If the timescale above which the power spectrum flattens to $\alpha \lsim 1$ ($\approx 2 \pi \tau$ in the DRW model, $\approx 1/f_{\rm low}$ in the BPL model) is shorter than the $\approx 1$ yr inter-season spacing of the data, phantom periods are rare. In such cases the inter-season variability is essentially white noise and this is unlikely to produce smooth undulations between seasons. On the other hand, if this timescale is considerably longer than the $\approx 10$ yr span of the observations, and the power spectrum remains steep far below the lowest observable frequencies (longest timescales), the variations will be dominated by smooth, quasi-linear trends that get weaker as $f_{\rm low}$ moves lower (in our models the total power in the power spectrum is constant, so power moves out of the observed band as $f_{\rm low}$ decreases). The chance of the variations being dominated by a succession of roughly equally spaced peaks (and/or troughs) is therefore reduced (unless one applies `detrending' to the data, which then increases the rate of phantom periods).

%%%%%%%%%%%%%%%%%%%%%%%%%%%%%%%%%%%%%%%%%%%%%%%%%%
\section{Seeing patterns in the noise}
\label{sect:noise}

These simulation experiments demonstrate that when trying to detect periodic signals from a large pool of red noise time series sampled like CRTS data, `phantom' periodicities will be found, and their periods tend to be near the longest allowed period ($1.5-2.5$ cycles over the available data, assuming obvious aliasing periods are ignored). This effect was previously discussed by \citet{Kozlowski2010} and \citet{MacLeod2010} from large surveys of quasar light curves. By contrast, genuinely periodic variations are most easily detected with periods $\sim 2$ yr due to the seasonal sampling of the data. We stress that the precise number of phantom periods we find should not be directly compared to the survey of \citetalias{Graham2015a}, \citetalias{Graham2015b}, or of \citet{Liu2015}, and \citet{Charisi2016} -- our detection criteria are different, and our simulation experiment focusses on above-average data quality (a bright, variable quasar). The key point is that light curves with a sinusoidal appearance are produced at a rate that is not insignificant, and this rate depends strongly on the power spectrum. Our lack of knowledge (or poor assumptions) about the true range of power spectral shapes for normal quasars translates to uncertainty about (or poor calibration of) the rate of phantom periods in quasar surveys.

That apparently sinusoidal time series are generated is not due to a problem with the analysis procedures, it is an intrinsic property of random processes with steep power spectra. Time series from red noise processes, which span timescales over which the power spectrum is steep ($\alpha \gsim 2$), will usually be dominated by smooth variations showing modulations occurring on timescales of order the length of the time series. Intermittent time sampling and low signal/noise of the data make it easier to mistake a phantom periodicity for a real one. Fundamentally they can only be distinguished by much longer time series spanning many cycles of the putative period; truly periodic processes will continue to oscillate while red noise processes are progressively less likely to show further oscillations. (An intermediate possibility, beyond the scope of this paper, is that of quasi-periodic oscillations (QPOs) which show drifts in period, phase or amplitude.)

%%%%%%%%%%%%%%%%%%%%%%%%%%%%%%%%%%%%%%%%%%%%%%%%%%
\section{The difficulty of selecting from large samples}

When searching for rare events in a very large survey of sources, it is particularly important to understand the false positive rate. Once the false positive probability per source is higher than the true incidence of the event in the survey population there will (on average) be more false than true detections. The periodic candidate identified by \citetalias{Graham2015a}, and the $111$ candidates identified by \citetalias{Graham2015b}, were selected from $\approx 250,000$ quasars. The characteristic timescales and other properties of quasar power spectra are still only poorly understood, but are likely to depend on the mass and other properties of the AGN \citep{McHardy2006, Kelly2009, Kelly2011, MacLeod2010}. The quasars in any large sample spanning a range of $z$ and $L$ will likely include a range of power spectral shapes. Those quasars with a high power density and steep spectra over the observed timescale range will be most likely to produce phantom periods, and a survey containing a few thousand such quasars should be expected to produce many phantom periods. 

\citetalias{Graham2015a} used DRW simulations to assess the significance of the PG 1302$-$102 detection. They performed two different tests, one was a simulation for each quasar in their survey, the other was an analysis of $1,000$ simulations of data like that of PG 1032$-$102. The apparent significance in the latter test only demonstrates that their simulations are not good at reproducing particular properties of the data, they do not demonstrate that a period has been detected. If quasar power spectra are steeper than the DRW model (as indicated by e.g. the \emph{Kepler} studies cited above) the simulation test based on DRW simulations could underestimate the number of false positives. By the same argument, it is meaningless to quote -- as \citetalias{Graham2015a} and \citet{Liu2015} do -- the detection significance of periodicities using periodogram statistics (including the Lomb-Scargle periodogram) that are calibrated against a white noise null hypothesis, when the alternative hypothesis is non-white noise. A small $p$-value in such cases simply rejects the white noise null hypothesis (already known to be false!), it does not necessarily support a periodic alternative. 

There are several other potential problems with the simulation test of \citetalias{Graham2015a}. One is that the PG 1302$-$102 was selected to be among the most periodic of $250,000$ quasars, hence there is a large `look elsewhere' effect\footnote{The `look elsewhere' effect, more generally known as the multiple comparison problem, occurs whenever an analysis includes many statistical tests or estimates. In the context of quasar surveys, where many quasar time series are each tested for periodicity, the chance of noise being mistaken for a period increases with the number of quasars in the survey (for a fixed test procedure). Allowing for greater flexibility in the tests being applied -- such as testing for transient periodicity or applying detrending to the data -- also increases the opportunities for false detections. Understanding or controlling the false discovery rate in large-scale surveys is an area of current research \citep{Algeri2016}.} for the number of quasars searched. Further, the wavelet method they used decomposes the data by time and frequency, increasing still more the `look elsewhere' effect. Precise calibration of the survey-wide false positive rate would require one to simulate the entire distribution of quasars many times over, accounting for the plausible range of aperiodic power spectra for each quasar, and with sufficient statistics to determine the probability of a false positive to an accuracy of $\lsim 10^{-5}$ per object. 

Another issue is the treatment of measurement errors in simulations. The CRTS photometric errors for PG 1302$-$102 are significantly overestimated; simulating random measurement errors that are larger than the measurement errors of the real data (as in \citealt{Graham2015a}) will lead to unrealistic simulations. In this case, that means too much `white noise' in the simulations, which then reduces the probability of the simulations producing strong, smooth modulations. If we repeat our simulations tests adding random measurement errors with a standard deviation equal to the CRTS pipeline error bars, the number of phantom periodicities drops by more than an order of magnitude. (Our selection procedure relies on obtaining a good fit, i.e. $\chi^2 \le dof$, which is much harder to achieve in the presence of increased white noise.)

%When considering candidates chosen from searches of large pools of data, it is difficult to interpret the results of model comparisons. In principle it is possible to formally compare stochastic models to periodic models (or combined models) in a formal sense, e.g. using Bayesian model comparison procedures. However, PG 1302$-$102 is not a randomly selected field quasar, it was chosen because it was among the strongest periodic candidates in a sample of nearly $250,000$. One should not be surprised if model selection often favours the periodic model when applied to data selected to be periodic from a large pool of data (even if those light curves are in fact realisations of a random process). In the particular case of PG 1302$-$102 we obtained $\chi^2 = 22.4$ using a DRW model with no periodic component. In a naive sense the stochastic model fits much better than the sinusoidal model (but, as shown by \citealt{Kozlowski2016}, one should not interpret the fit statistic so simply for stochastic models like this). 

%%%%%%%%%%%%%%%%%%%%%%%%%%%%%%%%%%%%%%%%%%%%%%%%%%
\section{Conclusions}
\label{sect:conc}

Fortunately, in particular cases such as PG 1302$-$102 \citepalias{Graham2015a} or PSO J334.2028+01.4075 \citep{Liu2015}, the issue of stochastic or nearly-periodic variations can be resolved by further observations. If more `cycles' of data -- ideally with a higher sampling rate and improved precision -- match the extrapolation of the current sinusoidal model, that would strongly support the periodic model. The more future cycles that remain coherent with the model (based on current data), the stronger the evidence for a true periodicity. If the light curve diverges from the model, and in an apparently random manner, that will be evidence against the periodic model. On the other hand a smooth and systematic period or phase drift could indicate the presence of an optical quasi-periodic oscillation (QPO), e.g. a strong resonance in the accretion flow not related to the orbit of a binary SMBH. 

In the short term, however, it is often more practical to `go wide' (light curves from many more targets) than to `go deep' (longer, better quality time series of individual targets), so it is particularly important to calibrate the false positive rate of few-cycle periodicities in irregular and noisy data. If true periodicities are rare -- as one might expect given than most quasars are powered by accretion onto a single BH and show stochastic variations, with only a small minority\footnote{\citetalias{Graham2015b} estimate that $1-10$ in every $1000$ field quasars (at $z < 1.0$) may be expected to harbour a short period SMBH binary system.} of spectroscopic quasars expected to be short period binary SMBHs -- a slight underestimate in the adopted value of the false positive rate could mean the false period detections outnumber the true detections\footnote{This is an example of the so-called `false positive paradox' of statistics.}. 

Improving our understanding of the range of quasar noise power spectra would improve any search for outliers in samples of quasar variability. Any true binary accreting black hole system likely undergoes stochastic variability in addition to periodic variations, so it is not yet clear that searches for the purely periodic signals are the best approach to finding these systems. We encourage further development of methods to identify periodic and mixed periodic/stochastic processes hidden among a range of stochastic processes, and to better identify the limitations of such methods when applied to sparsely sampled photometric data. In a preliminary investigation (applying the Bayesian analysis of section \ref{sect:bayes} on sinusoidal simulations from section \ref{sect:sims}) we found that even strictly sinusoidal variations were difficult to distinguish from a simple stochastic process when the number of cycles was $\lsim 2$, but relatively straightforward to distinguish with $\sim 5$ cycles. Further work needs to be done to uncover the fundamental limitations of distinguishing periodic and stochastic signals, given a particular sampling strategy. 

In 1978, Bill Press closed his article on \emph{Flicker noises in astronomy and elsewhere} with a note of caution about how easy it is for the eye-brain system to select `three-cycle' periodicities in random time series \citep{Press1978}. We might note here that, as we enter the era of `big-data' time domain surveys, one might do well to also regard with caution few-cycle periodicities selected by machine methods when they come from large samples of noisy time series. One should not be too pessimistic, however. As our understanding of the variability of different source populations improves, we will be better able to calibrate detection procedures and realise the potential of machine learning methods for mining the large time series compendia for rare, exotic behaviour.

\section*{Acknowledgements}
We thank an anonymous referee for a prompt and thoughtful report. SV acknowledges support from STFC consolidated grant ST/K001000/1. WNA acknowledges support from the European Union Seventh Framework Programme (FP7/2013-2017) under grant agreement n.312789, StrongGravity. MJM acknowledges support from an STFC Ernest Rutherford fellowship. DH acknowledges support by the Moore-Sloan Data Science Environment at NYU. This research made use of NASA's Astrophysics Data System. The CSS survey is funded by the National Aeronautics and Space Administration under Grant No. NNG05GF22G issued through the Science Mission Directorate Near-Earth Objects Observations Program.  The CRTS survey is supported by the U.S. National Science Foundation under grants AST-0909182 and AST-1313422. SV thanks 
Tom Maccarone for early discussions of some ideas in this paper, and C. Bailer-Jones for advice about his R code.

%%%%%%%%%%%%%%%%%%%%%%%%%%%%%%%%%%%%%%%%%%%%%%%%%%

%%%%%%%%%%%%%%%%%%%% REFERENCES %%%%%%%%%%%%%%%%%%

%%%%%%%%%%%%%%%%%%%%%%%%%%%%%%%%%%%%%%%%%%%%%%%%%%

%%%%%%%%%%%%%%%%% APPENDICES %%%%%%%%%%%%%%%%%%%%%

\appendix

\section{Supplementary information}

\subsection{More details on the simulations}
\label{sect:simdets}

For each stochastic simulation we generated a time series with $N = 2^{14}$ points and $400$ points per year, spanning $\approx 40$ years. The simulations are of linear, stationary, Gaussian processes with zero mean and a smooth (`broad-band noise') power spectrum, generated with the fast method of \citet[see also \citealt{Davies1987}]{Timmer1995}. Strictly, we simulate variations in the V-band magnitude, so the simulations naturally account for the lognormal distribution of (linear) flux \citep[see][]{Uttley2005}.

The power spectrum is normalised such that the expected total power (integral over all positive frequencies) is $7.5 \times 10^{-3}$ mag$^2$ (which translates to a fractional rms of $F_{\rm var} = \sigma_F / \langle F \rangle = 0.2$ for the linear fluxes). This is equivalent to $SF_{\infty} = 0.122$ mag in the notation of \citet{MacLeod2010}. To the simulated $40$ yr light curves we add an offset of $\langle V \rangle = 15.0$, ignore the first $10$ years of fake data (to mitigate `edge effects' in the simulation) and use linear interpolation between the regularly spaced ($dt = 21.915$ hr) fake data points to recover fluxes at  `observation' times with the same sampling pattern as the CRTS data for PG 1302$-$102. We then add independent random, Gaussian noise ($\mu = 0$, $\sigma = 0.015$ mag) to simulate observational noise. Finally, we truncate the magnitudes to two decimal places to represent the discretisation of the CRTS photometric data. Error bars are then assigned with $\sigma = 0.06$ mag, i.e. $4$ times larger than the random error (as in the real data; see section \ref{sect:obs}).

\citetalias{Graham2015b} include archival photometric data in addition to CRTS to extend the length of their light curve. The best sampled of these is the LINEAR data \citep{Sesar2011} which adds another two seasons of data prior to the start of the CRTS data. We have simulated this by taking the sampling pattern of the first three CRTS seasons and reversing it around the mid-point of the first CRTS season, to create new time points with realistic sampling extending $\approx 2$ seasons before the first CRTS season. The fake CRTS + LINEAR data contain $360$ data points.

For each periodic simulation we employ a similar procedure except the stationary Gaussian process is replaced by a sinusoid with a random phase (in the range $(-\pi, +\pi)$), an amplitude of $0.125$ mag (obtained by fitting the PG 1302$-$102 data; section \ref{sect:obs}), and a period drawn uniformly from the range $0.2 - 20$ yr, i.e. $\approx 0.5 - 50$ cycles over the $10.5$-year duration of the CRTS+LINEAR data. We then add the mean magnitude, apply the same time sampling pattern, include random errors, and discretize the magnitudes as above. 

\begin{figure}
	\includegraphics[width=\columnwidth]{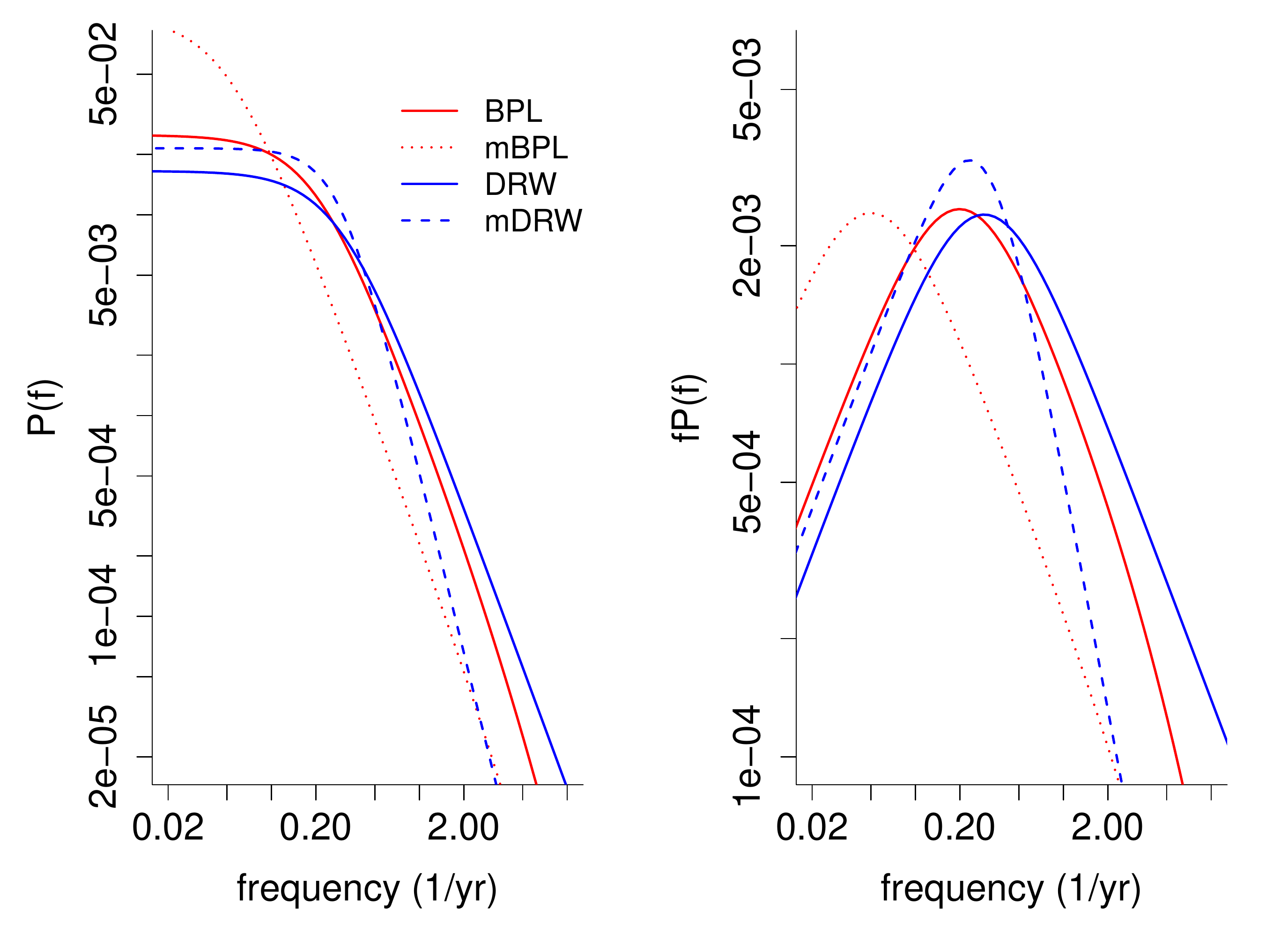}
	\vspace{-0.3cm}
    \caption{Power spectral models used for the simulations. The left panels show the power density $P(f)$, the right shows $f \times P(f)$ which better illustrates the power per decade in frequency. Frequency is in units of yr$^{-1}$. BPL is the (doubly) bending power law model with bend frequencies at $f_{\rm low} = 1/(5$ yr$)$ and $f_{\rm hi}=1/(50$ d$)$. The mBPL is a modified model with the low frequency bend moved down to $f_{\rm low} = 1/(20$ yr$)$. The DRW is the `damped random walk' model which bends to an index $-2$ above $f_{\rm bend} \sim (2 \pi \tau)^{-1}$. In this case $\tau = 200$ d (the mean from \citealt{MacLeod2010}). mDRW is a modified DRW model with a high frequency index of $-3$.}
    \label{fig:psds}
\end{figure}

%%%%%%%%%%%%%%%%%%%%%%%%%%%%%%%%%%%%%%%%%%%%%%%%%%

%%%%%%%%%%%%%%%%%%%%%%%%%%%%%%%%%%%%%%%%%%%%%%%%%%

% Don't change these lines
\bsp	% typesetting comment
\label{lastpage}

\begin{thebibliography}{99}
\bibitem[\protect\citeauthoryear{Algeri et al.}{2016}]{Algeri2016}
Algeri, S., van Dyk, D. A., Conrad, J., Anderson, B., 2016, Eur. J. Phys., submitted (arxiv:1602.03765)
\bibitem[\protect\citeauthoryear{Andrae, Kim \& Bailer-Jones}{2013}]{Andrae2013}
Andrae, R., Kim, D.-W., Bailer-Jones, C. A. L., 2013, A\&A, 554, A137
\bibitem[\protect\citeauthoryear{Bailer-Jones}{2012}]{Bailer-Jones2012}
Bailer-Jones, C. A. L., 2012, A\&A, 546, 89 
\bibitem[\protect\citeauthoryear{Begelman, Blandford \& Rees}{1980}]{Begelman1980}
Begelman M. C., Blandford, R. D., Rees, M. J., 1980, Nature, 287, 307
\bibitem[\protect\citeauthoryear{Charisi et al.}{2016}]{Charisi2016}
Charisi, M., Bartos, I., Haiman, Z., Price-Whelan, A. M., Graham, M. J., Bellm, E. C., Laher, R. R., Marka, S., 2016, MNRAS, submitted (arxiv:1604.01020) 
\bibitem[\protect\citeauthoryear{Davies \& Harte}{1987}]{Davies1987}
Davies, R. B., Harte, D. S., 1987, Biometrika, 74, 95–101 
\bibitem[\protect\citeauthoryear{Drake et al.}{2009}]{Drake2009}
Drake, A. J. et al., 2009, ApJ, 696, 870.
\bibitem[\protect\citeauthoryear{Edelson et al.}{2013}]{Edelson2013}
Edelson, R., Mushotzky, R., Vaughan, S., Scargle, J., Gandhi, P., Malkan, M., Baumgartner, W., 2013, ApJ, 766, 16
\bibitem[\protect\citeauthoryear{Edelson et al.}{2014}]{Edelson2014}
Edelson, R., Vaughan, S., Malkan, M., Kelly, B. C., Smith, K. L., Boyd, P. T., Mushotzky, R., 2014, ApJ, 795, 2
\bibitem[\protect\citeauthoryear{{Frank}, {King}, \& {Raine}}{2002}]{Frank2002}
Frank J., King A., Raine D. J., 2002, {Accretion Power in Astrophysics: Third Edition}. \newblock Cambridge University Press
\bibitem[\protect\citeauthoryear{Gierlinski et al.}{2008}]{Gierlinski2008}
Gierlinski, M., Middleton, M., Ward, M., Done, C., 2008, Nature, 455, 369
\bibitem[\protect\citeauthoryear{Graham et al.}{2015a}]{Graham2015a}
Graham M. J., et al. 2015a, Nature, 518, 74 
\bibitem[\protect\citeauthoryear{Graham et al.}{2015b}]{Graham2015b}
Graham M. J., et al., 2015b, MNRAS, 453, 1562 
%\bibitem[\protect\citeauthoryear{Gold et al.}{2014}]{Gold2014}
%Gold, R., Paschalidis, V., Etienne, Z. B., Shapiro, S. L., Pfeiffer, H. P., 2014, PhRvD, 89, 064060
\bibitem[\protect\citeauthoryear{Gonzalez-Martin \& Vaughan}{2012}]{Gonzalez-Martin2012}
Gonzalez-Martin, O, Vaughan S., 2012, A\&A, 544, 80
\bibitem[\protect\citeauthoryear{Goodman \& Weare}{2010}]{Goodman2010}
Goodman, J., Weare, J., 2010, Commun. Appl. Math. Comput. Sci., 5, 65
%\bibitem[\protect\citeauthoryear{Haehnelt \& Kauffmann}{2002}]{Haehnelt2002}
%Haehnelt, M. G., Kauffmann, G., 2002, MNRAS, 336, L61 
\bibitem[\protect\citeauthoryear{Haiman, Kocsis \& Menou}{2009a}]{Haiman2009a}
Haiman, Z., Kocsis, B., Menou, K., 2009a, ApJ, 700, 1952
\bibitem[\protect\citeauthoryear{Haiman}{2009b}]{Haiman2009b}
Haiman, Z., Kocsis, B., Menou, K., Lippai, Z., Frei, Z. 2009b, CQGra, 26, 094032
%\bibitem[\protect\citeauthoryear{Jaffe \& Backer}{2003}]{Jaffe2003}
%Jaffe, A. H., Backer, D. C., 2003, ApJ, 583, 616 
\bibitem[\protect\citeauthoryear{Kalamkar}{2016}]{Kalamkar2016}
Kalamkar, M., Casella, P., Uttley, P., O'Brien, K., Russell, D., Maccarone, T., van der Klis, M., Vincentelli, F., 2016, MNRAS, submitted (arXiv:1510.08907)
\bibitem[\protect\citeauthoryear{{Kelly}, {Bechtold} \& {Siemiginowska}}{2009}]{Kelly2009}
Kelly, B. C., Bechtold, J., Siemiginowska, A., 2009, ApJ, 698, 895 
\bibitem[\protect\citeauthoryear{Kelly, Sobolewska \& Siemiginowska}{2011}]{Kelly2011}
Kelly, B. C., Sobolewska, M., Siemiginowska, A., 2011, ApJ, 730, 52
\bibitem[\protect\citeauthoryear{Kelly et al.}{2014}]{Kelly2014}
Kelly, B. C., Becker, A. C., Sobolewska, M., Siemiginowska, A., Uttley, P., 2014, ApJ, 788, 33
\bibitem[\protect\citeauthoryear{Koz{\l}owski et al.}{2010}]{Kozlowski2010}
Koz{\l}owski, S., et al., 2010, ApJ, 708, 927
\bibitem[\protect\citeauthoryear{Koz{\l}owski}{2016}]{Kozlowski2016}
Koz{\l}owski, S., et al., 2016, MNRAS, submitted (arxiv:1604.01773)
\bibitem[\protect\citeauthoryear{Liu et al.}{2015}]{Liu2015}
Liu, T., et al., 2015, ApJ, 803, L16 
\bibitem[\protect\citeauthoryear{MacLeod et al.}{2010}]{MacLeod2010}
MacLeod, C. L., et al., 2010, ApJ, 721, 1014
\bibitem[\protect\citeauthoryear{McHardy et al.}{2006}]{McHardy2006}
McHardy, I. M., Koerding, E., Knigge, C., Uttley, P., Fender, R. P., 2006, Nature, 444, 730
\bibitem[\protect\citeauthoryear{Mushotzky et al.}{2011}]{Mushotzky2011}
Mushotzky, R. F., Edelson, R., Baumgartner, W., Gandhi, P., 2011, ApJ, 743, L12
\bibitem[\protect\citeauthoryear{Press}{1978}]{Press1978}
Press W. H. 1978, Comment. Astrophys., 7, 103 
\bibitem[\protect\citeauthoryear{Sesar et al.}{2011}]{Sesar2011}
Sesar B., et al., 2011, AJ, 142, 190
\bibitem[\protect\citeauthoryear{Summons et al.}{2007}]{Summons2007}
Summons, D. P., Arévalo, P., McHardy, I. M., Uttley, P., Bhaskar, A., 2007, MNRAS, 387, 649
\bibitem[\protect\citeauthoryear{Timmer \& K\"{o}nig}{1995}]{Timmer1995}
Timmer, J., K\"{o}nig, M., 1995, A\&A. 300, 707–710
\bibitem[\protect\citeauthoryear{an der Klis}{2006}]{vanderKlis2006}
van der Klis, M., 2006, in Compact stellar X-ray sources (W.Lewin \& M. van der Klis eds.), Cambridge University Press (Cambridge, UK), p. 39
\bibitem[\protect\citeauthoryear{Uttley, McHardy \& Vaughan}{2005}]{Uttley2005}
Uttley, P., McHardy, I. M., Vaughan, S., 2005, MNRAS, 359, 345
\bibitem[\protect\citeauthoryear{Vaughan \& Uttley}{2006}]{Vaughan2006}
Vaughan, S., Uttley, P., 2006, AdSpR, 38, 1405
\bibitem[\protect\citeauthoryear{Vaughan}{2013}]{Vaughan2013}
Vaughan, S., 2013, Phil. Trans. R. Soc. A, 371, 1984
\bibitem[\protect\citeauthoryear{Volonteri, Haardt \& Madau}{2003}]{Volonteri2003}
Volonteri, M., Haardt, F., Madau, P., 2003, ApJ, 582, 559 
\bibitem[\protect\citeauthoryear{Zheng et al.}{2016}]{Zheng2016}
Zheng, Z.-Y. Butler, N. R., Shen, Y., Jiang, L., Wang, J.X., Chen, X., Cuadra, J., 2016, ApJ, submitted (arxiv:1512.08730)
\bibitem[\protect\citeauthoryear{Zu et al.}{2013}]{Zu2013}
Zu, Y., Kochanek, C. S., Koz{\l}owski, S., Udalski, A., 2013, ApJ, 765, 106
\end{thebibliography}
\end{document}